\documentstyle[graphicx,preprint,prb,aps]{revtex}
\tightenlines
\begin{document}
\draft
\title{A vertical diatomic
artificial molecule in the intermediate coupling
regime in a parallel and perpendicular magnetic field }
\author{ F. Ancilotto$^1$,
D. G. Austing$^{2,5}$\thanks{corresponding author:
guy.austing@nrc.ca}
M. Barranco$^3$\thanks{corresponding author: manuel@ecm.ub.es},
R. Mayol$^3$,
K. Muraki$^{2}$,
M. Pi$^3$,
S. Sasaki$^{2}$,
S. Tarucha$^{2,4}$
}
\address{$^1$Istituto Nazionale per la Fisica della Materia and
Dipartimento di Fisica, Universit\`a di Padova, I-35131 Padova,
Italy}
\address{$^2$NTT Basic Research Laboratories, 3-1
Morinosato Wakamiya, Atsugi, Kanagawa, 243-0198, Japan}
\address{$^3$Departament ECM,
Facultat de F\'{\i}sica, Universitat de Barcelona, E-08028 Barcelona,
Spain}
\address{$^4$Departament of Physics and ERATO Mesoscopic Correlation
Project, University of Tokyo, 7-3-1 Hongo, Bunkyo-ku, Tokyo, 113-0033,
Japan}
\address{$^5$Institute for Microstructural Sciences M23A, National
Research Council, Montreal Road, Ottawa, Ontario K1A 0R6, Canada}
\date{\today}

\maketitle

\begin{abstract}

We present experimental results for the ground state electrochemical
potentials of a few electron semiconductor artificial molecule
made by vertically coupling two quantum dots,
in the intermediate coupling regime, in perpendicular
and parallel magnetic fields up to $B \sim$ 5 T. We perform
a quantitative analysis based on local-spin density functional
theory. The agreement between theoretical and experimental results is
good, and the phase transitions are well reproduced.
\narrowtext

\end{abstract}


\section{Introduction}

Semiconductor quantum dots (QD's) are widely regarded as artificial
atoms with properties analogous to those of `natural' atoms.
Furthermore, systems composed of two QD's, `artificial'
quantum molecules (QM's), coupled either laterally or vertically, have
recently been investigated experimentally\cite{EXPBLOCK} and
theoretically.
\cite{lateor,Ron99,Par00,Tok00,Pi01}
Transistors incorporating
QM's\cite{Aus98} made by vertically coupling two well defined and highly
symmetric QD's\cite{Tar96} are ideally suited
to study QM properties.
We recently reported the addition energy spectra at zero magnetic field
for such QM's as a
function of interdot coupling strength.\cite{PiPRL01}

In this work we present experimental and theoretical
ground state electrochemical potentials for a diatomic QM
in the intermediate coupling regime corresponding to an interdot
distance, $b=3.2$ nm, for magnetic fields $(B)$ up to about 5 T. We
assume here that the quantum mechanical coupling is sufficiently strong
that the QM can be regarded as a symmetric `homonuclear' diatomic QM.
\cite{PiPRL01}
We consider two different configurations, one corresponding
to an applied magnetic field parallel $(B_{\parallel})$ to
the drain current $I_{\rm d}$ flowing through the constituent
QD's, and another corresponding to an applied magnetic field
perpendicular $(B_{\perp})$ to $I_{\rm d}$ (see Fig. \ref{fig1}).
The latter has received relatively little attention.
\cite{Tok00,Bur00,Sas98,Sas02}
We note that the QM physics we discuss in both magnetic field
configurations is particularly
relevant to the subject of solid-state quantum computing.\cite{Bur00}

The interpretation of the experimental results here is based on the
application of  local-spin density-functional theory (LSDFT).
\cite{Pi01,Per81,DFT/SD}
In the $B_{\parallel}$ case it follows the development
of the method thoroughly described in Ref. \onlinecite{Pi01}, which
includes finite thickness effects of the dots, and uses a relaxation
method to solve the partial differential equations arising from a
high order discretization of the Kohn-Sham (KS) equations on a spatial
mesh in cylindrical coordinates (axial symmetry  is assumed).
To describe the less-common $B_{\perp}$ case, a  three-dimensional (3D)
LSDFT code has been developed to handle configurations without any spatial
symmetry.

This work is organized as follows. In Sect. II we describe the
experimental setup; in Sect. III we outline the method used to
implement LSDFT in our axially symmetric
QM system, and discuss the fully 3D configurations; and in Sec.
IV we give the experimental and theoretical results. The interpretation
of these results, and a short summary is presented in Sect. V.

\section{Experiment}

The molecules we study are formed by coupling, quantum mechanically
and electrostatically, two QD's which individually display clear
atomic-like
features.
\cite{Aus98,Tar96}
For the materials we typically use, the energy splitting between the
bonding and anti-bonding sets of single particle (s.p.) molecular
states,
$\Delta_{\rm SAS}$, can be varied from about 4.5 meV (strong coupling)
to about 0.1 meV (weak coupling).\cite{Pi01,Aus98}
In this paper, $b$, the thickness of the central barrier separating the
two dots is fixed at 3.2 nm ($\Delta_{\rm SAS}\sim 3$ meV). Because this
corresponds to intermediate coupling, we can reasonably
neglect a small mismatch
(of energy $<\Delta_{\rm SAS}$) between the two dots, i.e. the QM
is
assumed to be symmetric `homonuclear'.
\cite{PiPRL01}
Figure \ref{fig1} shows (a) a schematic section of a sub-micron
circular mesa, diameter $D$, containing two vertically coupled QD's,
and (b) a scanning electron micrograph of a typical mesa after gate
metal deposition.  The starting material, a special triple barrier
resonant tunneling structure, and the processing recipe are described
elsewhere.
\cite{Aus98,SST,Ama01} Drain current $I_{\rm d}$ flows through the two
QD's between the substrate contact and grounded top contact in response
to voltage $V_{\rm d}$ applied to the substrate, and gate voltage
$V_{\rm g}$ on the single surrounding gate. The one structure we
describe here ($D \sim$ 0.5 microns) is cooled to about 300 mK or less.

\section{Theory}

To analyze the experiments we have modeled the QM by two
identical QD's stacked in the $z$ direction (parallel to $I_{\rm d}$).
In this direction the QM is confined by two identical quantum wells
of width $w$ and depth $V_0$ separated by distance $b$= 3.2 nm. We
have taken $V_0=225$ meV and $w=12$ nm, which are also appropriate
for the actual experimental device.
To improve on the convergence of the 3D calculations, the somewhat
ideal sharp double-well profile has been slightly rounded-off, as
shown in Fig. \ref{fig2}. Given that the energy profiles of real
structures are never abruptly sharp, this rounding-off is actually
not unrealistic.

In the remaining two directions perpendicular to $z$ the QM is confined
by a harmonic oscillator potential $V_h=m \omega^2 (x^2+y^2)/2$ of
fixed strength $\hbar \omega= 4.42$ meV. This  lateral confinement
energy has been determined for $N=6$ electrons using a general
law\cite{Aus02}
that quantitatively describes the phases of QM's in the strong,
intermediate and weak coupling regimes as a function of $B_{\parallel}$
for a number of
electrons, $N$, between 12 and 36. Lacking a better prescription at
smaller $N$,\cite{nota} $\omega$ has been kept fixed for all $N$
analyzed here, and in this sense we present a parameter-free
calculation. In the following we will denote by $V_{cf}(x,y,z)$
the total confining potential obtained by adding the double well
profile to the harmonic
oscillator potential, $V_h$. We stress that $V_{cf}(x,y,z)$ is axially
symmetric around the $z$ axis.

As it is well known, within LSDFT the ground state (g.s.) of the
system is
obtained by solving the Kohn-Sham (KS) equations. In the $B_{\parallel}$
case the problem is greatly simplified by explicitly using the axial
symmetry of the system. The additional terms in the KS equations due
to the
presence of an arbitrary magnetic field are given below. The inclusion
of these terms crucially does not break the axial symmetry of the KS
Hamiltonian in the $B_{\parallel}$ case.

In the symmetric gauge the vector
potential $\vec{A}(\vec{r}\,)$ corresponding to a constant magnetic
field $\vec{B}$ is written as
$\vec{A} =(\vec{B} \wedge \vec{r}\,)/2$, and its contribution to the
KS Hamiltonian is

\begin{equation}
{\cal H}_m = \frac{e \hbar}{2 m c}\, \vec{B} \cdot \vec{L} +
\frac{e^2}{2 m c^2} \vec{A}^{\,2} + g_s^* \,\mu_B \,\vec{B} \cdot
\vec{S}
\,\,\, ,
\label{eq1}
\end{equation}
where $g_s^*$ is the effective gyromagnetic factor,
$\vec{L}$ and $\vec{S}$ respectively are the orbital and spin 
angular momentum
operators, and $\mu_B$ is the Bohr magneton. Writing
$\vec{B} = B ( {\rm sin} \theta_B, 0, {\rm cos} \theta_B)$
and introducing the cyclotron frequency $\omega_c = e B/m c$, it can
be easily checked that ${\cal H}_m = {\cal H}_{m_R} + i {\cal H}_{m_I}$,
with

\begin{eqnarray}
{\cal H}_{m_R} &=& \frac{1}{8}\,m \,\omega_c^2 \,[ x^2 {\rm cos}^2 \theta_B
+ y^2 + z^2 {\rm sin}^2 \theta_B - 2 \,xz\, {\rm sin}\,
\theta_B\, {\rm cos}\, \theta_B]
+ \frac{1}{2} \,g_s^* \,\mu_B \,\eta_{\sigma} B
\nonumber
\\
& &
\label{eq2}
\\
{\cal H}_{m_I} &=& -\frac{1}{2}\,\hbar \omega_c\,
\left[{\rm sin}\, \theta_B \,\left( y \frac{\partial}{\partial z}
 - z \frac{\partial}{\partial y}\right) +
{\rm cos}\, \theta_B \,\left( x \frac{\partial}{\partial y}
 - y \frac{\partial}{\partial x}\right)\right] \,\,\, ,
\nonumber
\end{eqnarray}
where $\eta_{\sigma}$=$+1(-1)$ for $\sigma$=$\uparrow$$(\downarrow)$
with respect of the direction of the applied magnetic field.

We have used effective atomic units
$\hbar=e^2/\epsilon=m=$1, where
$\epsilon$ is the dielectric constant,
and $m$  the electron effective mass. In units of the bare
electron  mass $m_e$, $ m = m^* m_e$.
In this system, the length unit is the effective
Bohr radius $a_0^* = a_0\epsilon/m^*$ with $a_0=\hbar^2/m_e e^2$,
and the energy unit is the effective Hartree $H^* = H  m^*/\epsilon^2$.
For a QD in GaAs,  we take the following values:
$g_s^*= -0.44$, $\epsilon$ = 12.4, and  $m^*$ = 0.067.
This yields $a^*_0 \sim$ 97.94 ${\rm \AA}$ and $H^*\sim 11.86$
meV. From now on we will write the equations in these units.

Equation (\ref{eq2}) reduces to the $B_{\parallel}$ case when
$\theta_B =0$, and to the $B_{\perp}$ case when
$\theta_B =\pi/2$. In the former, since
$( x \frac{\partial}{\partial y}
 - y \frac{\partial}{\partial x})$ is proportional to $L_z$, the
problem remains axially symmetric.
A detailed description of
how the KS equations have been solved in this geometry
can be found in Ref. \onlinecite{Pi01}.

In 3D the KS equations read

\begin{eqnarray}
& & \left[-\frac{1}{2} \left( \frac{\partial^2}{\partial x^2}
+ \frac{\partial^2}{\partial y^2}
+ \frac{\partial^2}{\partial z^2} \right)
+ V_{cf}(x,y,z) \right.
\nonumber
\\
& &
\label{eq3}
\\
&+& \left. V^H + V^{xc} + W^{xc}\,\eta_{\sigma}
+ {\cal H}_m \right]
\Psi_{\sigma}(x,y,z) =
\epsilon_{\sigma} \Psi_{\sigma}(x,y,z) \,\, .
\nonumber
\end{eqnarray}
The expression in the brackets is the KS Hamiltonian ${\cal H}_{KS}$,
and
$V^H(x,y,z)$ is the direct Coulomb potential. $V^{xc}={\delta
{\cal E}_{xc}(n,m)/\delta n}\vert_{g.s.}$, and
$W^{xc}={\delta {\cal E}_{xc}(n,m)/\delta m}\vert_{g.s.}$
are, respectively, the variation of the exchange-correlation
energy density ${\cal E}_{xc}(n,m)$ in terms of the electron
g.s. density $n(x,y,z)$, and of the local spin magnetization
$m(x,y,z)\equiv n^{\uparrow}(x,y,z)-n^{\downarrow}(x,y,z)$.
The exchange-correlation energy has been taken from
Perdew and Zunger,\cite{Per81} and
${\cal E}_{xc}(n,m)$
has been constructed as indicated in Ref. \onlinecite{Pi01}.
It is worth noticing that if $B\ne0$  then the s.p.
wave functions $\Psi_{\sigma}(x,y,z)$ are complex, with their real and
imaginary parts being coupled by ${\cal H}_m$.

The KS and Poisson equations are solved on a 3D mesh
after discretizing them using 7-point formulas,
and using a two-grid version of the one-way multigrid method
described in Ref. \onlinecite{Lee00}.
The Poisson equation is solved using a first order
relaxation scheme.\cite{Pre92}
The required value of the Coulomb potential
at the mesh boundary is obtained by a standard multipole
expansion up to eighth order.
The KS equations are solved using an imaginary
time method, involving the third-order expansion
of the forward solution of the imaginary time diffusion 
equation\cite{Pre92}

\begin{equation}
\frac{\partial \Psi}{\partial \tau} = - ({\cal H}_{KS} - \epsilon)\Psi
\label{eq4}
\end{equation}
i.e.,

\begin{eqnarray}
& & \Psi(\tau+\delta\tau) - \Psi(\tau) \equiv \Delta \Psi(\tau)
= - \delta\tau ({\cal H}_{KS}-\epsilon) \Psi(\tau)
\nonumber
\\
& &
\label{eq5}
\\
&+& \frac{\delta\tau^2}{2} ({\cal H}_{KS}-\epsilon)^2 \Psi(\tau) -
\frac{\delta\tau^3}{6} ({\cal H}_{KS}-\epsilon)^3 \Psi(\tau)
\,\,\, ,
\nonumber
\end{eqnarray}
where $\epsilon=\langle\Psi(\tau)|{\cal H}_{KS}|\Psi(\tau)\rangle$.
To further accelerate the self-consistent solution of both the KS and
Poisson equations, we use the preconditioning smoothing operation
described in Ref.\onlinecite{Hos95}. In the KS case, this means that
$\Delta \Psi(\tau)$ has been smoothed as proposed in this reference.
The perfomance of the code has been further improved by adding a
`viscosity term', i.e., Eq. (\ref{eq5}) has been changed into

\begin{equation}
\Psi(\tau + \delta\tau) - \Psi(\tau) =
\Delta \Psi(\tau) +\alpha_V [\Psi(\tau) - \Psi(\tau - \delta \tau)]
\,\,\, .
\label{eq5a}
\end{equation}
A viscosity term has also been included in the solution of the
Poisson equation. We have used a $45 \times 45 \times 67$ mesh with
spatial steps
$\Delta x = \Delta y = 5.67 $ nm, and $\Delta z = 0.89$ nm.
The large asymmetry between the spatial meshes is motivated by
the sharpness of the confining potential in the $z$ direction.
The heuristic viscosity parameter $\alpha_V$ is fixed to a value
of 0.8, and the time step $\delta \tau$  to the
value of $(\Delta z)^2$ in effective atomic units. The stability
of our results against the increase of the number of mesh points
has been checked, and at $B=0$  we have used the results of the
axially
symmetric code to test the results obtained with the 3D code.

The accuracy of LSDFT for the $B$ values of interest has been
assessed by comparing the results for a single QD
with those obtained using the current spin-density functional theory
(CSDFT),\cite{Fer94,Pi98,Ste98,Rei99,Par01} which in principle is
better suited to high magnetic fields than LSDFT.
Since CSDFT is a two dimensional (2D) theory,\cite{Fer94}
we have also compared our LSDFT results with those obtained using the
2D-LSDFT which is implicit in any implementation of CSDFT, in particular
see that of Ref. \onlinecite{Pi98}.
The low and high field borders
of the maximum-density droplet (MDD) phase
using strictly 2D-LSDFT and CSDFT, as described in
Ref. \onlinecite{Pi98}, have been obtained for QD's with $N=20$, 28 and
36 electrons laterally confined by a harmonic oscillator potential of
energy $\hbar \omega= 7.6 \,N^{-1/4}$ (meV). Such
parameterization of the confining potential within LSDFT reproduces the
experimental MDD. \cite{Oos99}
The results are shown in Table \ref{Tab1}, together with the values
obtained by using the present 3D-LSDFT for the same lateral confining
potential. From Table \ref{Tab1} we can see that the overall agreement
between the three calculations is clearly good, and thus we can
confidently use LSDFT in the present calculations.

\section{Results and discussion}

The experimental ground state electrochemical potentials for
$N=3$ to 6, as a function of $B$,
are shown in Fig. \ref{fig3} for (a) the parallel case, and (b) the
perpendicular case. What is actually shown is the $B$-field dependence
of the 3rd, 4th, 5th and 6th current (Coulomb oscillation) peaks
measured by sweeping $V_{\rm g}$ in the linear conductance regime
for a small $V_{\rm d} \simeq 0.1$ mV.
It is clearly evident that the dependencies for parallel and
perpendicular cases are very different -- in particular the former is
stronger than the latter. We now attempt to explain the general
appearance in both cases, and in particular the features marked by the
different symbols, by using the computational methods described in the
previous section.

We first address the $B_{\parallel}$ case.
As in  single QD's, at low $B$-fields,
upward kinks (cusps) in the experimental
$N$ electron g.s. QM electrochemical potentials as a function of $B$
are interpreted as changes in the $N$ electron g.s. configuration of
the QM which arise from s.p. level crossings.\cite{Tar96,Oos99,Leo97}
We have plotted in Fig. \ref{fig4} the calculated
g.s. electrochemical potential $\mu(N)$, defined as
\begin{equation}
\mu(N) = U(N) - U(N-1) \,\,\, ,
\label{eq6}
\end{equation}
where $U$ is the total energy of the $N$ electron QM g.s.,
as a function of $B_{\parallel}$ up to $N=6$.
To label the  g.s. configurations we have used the usual
notation of
molecular physics for s.p. electronic orbitals.\cite{box}
Upper case Greek letters are used for the total orbital angular
momentum. We have also used the adapted version\cite{Ron99} of
the ordinary spectroscopic notation $^{2S+1}L_{g,u}^{\pm}$
with $S$ being the total $|S_z|$, and $L$ being the  total $|L_z|$.
The superscript $+(-)$ refers to even (odd)
states under reflection with respect to the $z=0$ plane bisecting the
QM.
Even states are bonding (symmetric) states, and odd states are
anti-bonding (anti-symmetric) states. The subscript $g(u)$ refers to
positive
(negative) parity states. All these are good quantum numbers in the
$B_{\parallel}$ case and can be used to label the different g.s.'s
(`phases').
Following Refs. \onlinecite{Par00,Par01}
we have also calculated the `isospin' quantum number
(the `bonding number' in molecular physics)
defined as $I_z=(N_B-N_{AB})/2$,
with $N_{B(AB)}$ being the number of electrons in bonding
(anti-bonding) s.p. states.
This is an exact quantum number for homonuclear QM's in a parallel
magnetic field.\cite{noteloc}

Given the complexity of real vertical QM structures and the challenge in
modelling them,\cite{Ron99,Par00,PiPRL01,Par01}
a comparison between Figs. \ref{fig3}(a) and \ref{fig4} reveals a
rather good agreement between theory and experiment. As a guide,
and consistent with the calculated states and the observed $B_{\parallel}$
dependence,
we indicate in Fig. \ref{fig3}(a)
in simple box style\cite{box} the dominant g.s. configurations at or
near $B=0$, and others at higher field which are stable over a relatively
wide range of $B_{\parallel}$. Up and down arrows indicate spin-up and
spin-down electrons, and black (grey) arrows represent electrons in
bonding (anti-bonding)  s.p.  states.\cite{Leo97,box}
For $N=3$, 5, 6, near $B=0$, because the g.s.'s are close to each other,
i.e. stable over a fairly narrow range, we show two configurations
which in practice  are
hard to resolve. Some of these involve the population of the
lowest anti-bonding state with a single electron, so isospin is
non-maximal. Above $B=1$ T, however, all the anti-bonding states are
depopulated so isospin is maximal ($I_z=N/2$), and filling of the
QM resembles that of a single QD. The identifiable g.s. transitions
in Figs. \ref{fig3}(a) are marked by black triangles. As expected,
most appear as upward kinks. A couple, see first kinks for $N=5$ and 6,
appear as "downward" kinks because of the g.s. transitions which occur
at almost the same $B_{\parallel}$ in $N=4$ and 5 respectively.

Looking further at other details in Fig. \ref{fig4}, for $N=2$,
the singlet-triplet transition occurs at about 4.6 T
which is close to the experimental value\cite{Ama01} of $\sim 4.2$ T
(not shown).
We have found from the calculations an MDD configuration made of
electrons filling just
bonding s.p. states (MDD$_{\rm B}$),
which has a total angular momentum $L_z = N(N-1)/2$,
and extends from $\sim 4.9$ to $\sim 9.5$ T for $N=3$,
from $\sim 5.1$ to $\sim 9.0$ T for $N=4$,
from $\sim 5.4$ to $\sim 8.8$ T for $N=5$, and
from $\sim 5.6$ to $\sim 8.3$ T for $N=6$.
These results are at variance with those of Ref.
\onlinecite{Par01}, where an MDD$_{\rm B}$ g.s. was
found for $N=3$, but not for larger values of $N$.
The reason of this discrepancy may be attributed
either to the strictly 2D model used in their
calculation to represent the constituent QD's, or
more likely to their particular implementation of
CSDFT.\cite{nota0} A quantitative description of the spin phases of
QM's in different coupling regimes, along with the experimental data,
will be presented elsewhere.\cite{Aus02}

In Fig. \ref{fig4} we can see that for the larger $N$ values studied here,
the increase in angular momentum of the QM g.s. as it evolves from $B=0$
towards the MDD$_{\rm B}$ is accompanied by two isospin
flips\cite{Par01} caused by electrons jumping from anti-bonding to
bonding states and vice-versa.
Phase transitions  from $-$ to $+$ g.s.'s involve
$\Delta I_z=1$ flips, whereas those
from $+$ to $-$ g.s.'s involve $\Delta I_z=-1$ flips, and both are clearly
seen in Fig. \ref{fig4} for $N=5$ and 6.
Interestingly, they only happen for $B_{\parallel}<2$ T.
We can see that after reaching the $\nu_B=2$
g.s. (i.e., a filling factor two QM state made of just bonding
s.p. states),
which corresponds to the $^1\Delta_g^+$ phase for $N=4$,
to the $^2\Gamma_g^+$ phase for $N=5$,
and to the $^1I_g^+$ phase for $N=6$, only bonding s.p. states are
occupied, and as a consequence the QM reaches the MDD$_{\rm B}$ state
in a similar way to how a single QD reaches the MDD state, namely by
populating bonding s.p. states of higher and higher s.p. orbital
angular momentum $l$ values.\cite{note1}
In general, these isospin flips can produce a complex pattern in the
s.p. spectrum as a function of $B_{\parallel}$. As an example of this
complexity, we present in Fig. \ref{fig5} the s.p. levels for $N=6$
as a function of $l$ for
different $B_{\parallel}$ values.
It can be seen in this figure that as $B$ increases, the
QM undergoes isospin flips. Firstly, the $l=0$$\uparrow$ anti-bonding
s.p. state becomes occupied, as shown in the panels corresponding
to $B=0.5$ and 1.2 T. After another isospin flip caused by the
depopulation of the same s.p. state,
the QM reaches the $\nu_B=2$ phase corresponding to the  $^1I^+_g$
configuration ($B=3$ T panel). From this phase
on, the  spin polarization steadily increases until the QM
reaches the MDD$_B$ phase ($B=6$ T panel).

In the $B_{\perp}$ case, crucially, within LSDFT, the only
good quantum number is the spin projection {\em along} the direction of
the applied magnetic field, which we call $s_{\perp}$, and the g.s.
electrochemical potentials as a function of $B_{\perp}$ are expected
to be much smoother than in the $B_{\parallel}$ case. This situation
\cite{Tok00,Bur00,Sas98,Sas02}
unlike the $B_{\parallel}$ case, lacks an analytical solution even for
the
case of non-interacting electrons.
We show in Fig. \ref{fig6} the calculated non-interacting s.p.
spectrum as a function of $B_{\perp}$.
At $B=0$  the energy difference between bonding and anti-bonding
$l=0$ s.p. states is just $\Delta_{SAS}$ (likewise for
the $l=1$ states). The energy difference
between $l=1$ and 0 bonding (or anti-bonding) states is just
$\hbar\,\omega$. Similar results have been reported elsewhere.
\cite{Tok00,Sas02} Note that the effective value of $\Delta_{SAS}$ is
reduced as $B_{\perp}$ is increased because the tunneling electron
trajectories are increased in the barriers.
The small splitting between spin up and down states that originate
from a common s.p. state with well defined orbital angular momentum
at $B=0$ are due to the Zeeman term.
Increasing $B_{\perp}$ has also an effect on the quantum mechanical
 coupling of
the QD's forming the QM, as the already sizeable electron density
in the middle of the central barrier increases further (see $n(z)$ in
Fig. \ref{fig2}) which magnifies exchange-correlation effects between both
QD's. The compression of $n(z)$ in each well as  $B_{\perp}$ increases shown in
Fig. \ref{fig2} can be understood from  Eq. (\ref{eq2}): note  that when
$\theta_B=\pi/2$, the first term in ${\cal H}_{m_R}$ becomes
$ \omega_c^2 \,(y^2 + z^2)/8$.
This also produces an upward shift of the
s.p. spectrum of the QM as a whole
(the horizontal dashed line in Fig. \ref{fig2} lies above the
horizontal solid line in the same figure).

The even (bonding) or odd (anti-bonding) character
of the s.p. levels defining a QM state is strictly lost when a magnetic
field perpendicular to $I_{\rm d}$ is present. Intriguingly, however,
the bonding/anti-bonding character present at $B=0$  is sometimes
retained to a large degree by the s.p. states at finite
$B_{\perp}$ values. We have indeed found that the expectation value
of the $z\rightarrow -z$ reflection operator

\begin{equation}
\langle\Pi_z\rangle = \int d\vec{r} \, \Psi^*(\vec{r}\,)
\Pi_z \Psi(\vec{r}\,)=
 \int d \vec{r} \, \Psi^*(x, y, z)
 \, \Psi(x, y, -z)
\label{eq7}
\end{equation}
is very close to $\pm 1$, as it should be for bona fide
bonding/anti-bonding states,  in many cases even for relatively large
values of $B_{\perp}$.

As an example of this, we show in Fig.
\ref{fig7} the energies of the occupied s.p. states
as a function of $B_{\perp}$ for 
$N=5$ and 6. Solid triangles represent `quasi-bonding' states with
$\langle\Pi_z\rangle \geq 0.95$. Note that at 0 T
for $N=5$ the s.p. bonding state
at $\epsilon\sim 48.2$ meV is two fold degenerate, and likewise for 
two of the $N=6$ s.p. bonding states at $\epsilon\sim 52$ meV.
Open triangles represent
`quasi-anti-bonding' states with $\langle\Pi_z\rangle \leq -0.95$.
Actually, there is only one such occupied anti-bonding s.p. state
for $N=5$ at $B=0$, and none for $N=6$. All other open symbols
(circles and squares) correspond to s.p. states with negative
$\langle\Pi_z\rangle$ values larger than -0.95, i.e., cannot really 
be regarded even as `quasi-anti-bonding' states.

The figure also shows that states that evolve
from $l=0$ s.p. states at $B=0$  retain a quasi-bonding character up to
quite high values of $B_{\perp}$ (at least up to 5 T), 
whereas other states, that at $B=0$
are $l=1$ s.p. states, do not.
The quasi-bonding robustness of the lower-lying
s.p. states may be due to the
small effect that the applied magnetic field has on states that
are $l=0$ s.p. states at $B=0$.
The $B_{\perp}$ evolution of what at 0 T are the $2p$-states is rather
similar for $N=5$ and 6 with a change from solid to open symbols near
$B_{\perp}=4$ T.
Interestingly, in spite of the lack of any spatial
symmetry in the system when a perpendicular field is applied,
the s.p. levels are still clearly distributed into shells
as in the non-interacting case.\cite{Tok00,Sas02}
Notice also the different splitting between $\uparrow$ and $\downarrow$
states. For saturated (zero) spin ($N=6$ case),
this is essentially due to the
small Zeeman term, whereas for non-saturated spin ($N=5$ case)
the splitting is mostly due\cite{Pi98} to the spin-dependent
part of the exchange-correlation energy, $W^{xc}$ term in Eq.
(\ref{eq3}), and this effect is larger the higher the value of the g.s.
spin.
This explains the sizeable splitting between the two lower
lying s.p. levels for $N=5$ up to $B_{\perp}\sim 0.5$ T
and the splitting of all the s.p. levels for $N=6$ above
$B_{\perp}\sim 3.5$ T (see also Fig. \ref{fig8}).

The calculated g.s. electrochemical potentials are shown in Fig.
\ref{fig8} as a function of $B_{\perp}$.
Comparing with Fig. \ref{fig3}(b) it can be seen
that the agreement with experiment is good for
$3 \le N \le 6$. We have indicated
the value of the total $S_{\perp}$ for all the relevant g.s. phases. In
the $B=0$
to 5 T range, there are some $B_{\perp}$ induced changes in
$S_{\perp}$, and these give rise to upward kinks (also marked
in Fig. \ref{fig3}(b) by  solid down triangles). Some downward
kinks, identified by  vertical arrows
 in the $N=5$ and 6 g.s. electrochemical
potentials do not correspond to changes in the $N$-electron
$S_{\perp}$. They are associated with s.p. level crossings
between s.p. states of different
$\langle\Pi_z\rangle$ value of the $N-1$ electron system. This is the
case for $N=5$ at $B_{\perp} \sim 1.5$ T and $N=6$ at $B_{\perp} \sim
0.75$ T, as can be seen in Fig. \ref{fig8}
(also marked in Fig. \ref{fig3}(b) by solid circles).
Because of the lack of
spatial symmetry in the system, we do not, in general, attempt to
identify the (dominant) g.s. configurations. The configurations shown
in simple box style in Fig. 3 (b) are the dominant g.s.
configurations at 0 T and they are expected to remain so for small
values of $B_{\perp}$.
The singlet-triplet transition for $N=2$ in Fig. 8 appears at
$\sim 4.7$
T, a value compatible with that found in the $B_{\parallel}$ case.
The $B_{\perp}$ induced singlet-triplet transition in the
experimental data is discussed elsewhere.\cite{Sas02} It can
also be seen that for $N=4$ Hund's first rule like filling occurs for
$B_{\perp}<$ 1.5 T, even if the g.s. configuration is not strictly
axially symmetric. Nonetheless, for $N=6$  at $B=0$
we have found an axially symmetric configuration corresponding to a
2D harmonic oscillator shell like filling.

\section{Summary}

We have thoroughly discussed the ground state electrochemical
potentials
of a few-electron semiconductor artificial QM in the intermediate
coupling regime. A detailed comparison between experimental data
and LSDFT calculations shows overall a good agreement for
both parallel and perpendicular magnetic fields. The agreement
is even more remarkable since
the frequency $\omega$ of the lateral confining potential
has not been used here as a fitting parameter, but rather it has been
derived from a law strictly valid for larger values of $N$. Had
we used an even smaller value of $\omega$, the agreement would have
been even better.

Any sensible comparison with the results of
other calculations\cite{Ron99,Par01}
and with the experimental data should consider the strong influence
of the confinement on the actual g.s. QM phases. Large
$\omega$ values obviously favor the occupation of anti-bonding
states. Consequently, decreasing $\omega$ might `wash-out' phase
transitions involving anti-bonding states.
For example, see the $^2\Pi^+_u \rightarrow \,^2\Delta^-_u$ transition
for $N=5$ or the $^1\Sigma^+_g \rightarrow \,^3\Pi^-_g$ transition for
$N=6$ in Fig 4. We have checked that this is indeed the case when the
$N$ dependent confining potential
$\hbar \omega = 5.78 \,N^{-1/4}$ (meV)
of Refs. \onlinecite{Ron99,Par01} is used.
The same is true for the
$^2\Sigma^-_u \rightarrow \,^2\Pi^+_u$ transition for
$N=3$ (these two states are practically degenerate at $B=0$).
Thus, a comparison between theory and experiment may help to find
accurate and realistic values for the effective lateral confining
potential.

Analysis of the $B_{\perp}$ case has shown that the $N=2$ singlet-triplet
transition sensibly occurs at a similar $B$ value to that in the
$B_{\parallel}$ case. In spite of the absence of the strong
Landau quantization inherent to the $B_{\parallel}$ situation,
the s.p. levels are, to a large extent, still distributed into shells.
We have also found that spin effects arising from the
spin dependence of the
exchange-correlation energy can dominate those caused by
the small Zeeman term as is also the case for a single QD.

\section*{Acknowledgments}
This work has been performed under grants PB98-1247  from
DGESIC and 2001SGR-00064 from Generalitat of Catalunya, and
by the Specially Promoted Research, Grant-in-aid for Scientific
Research, and by DARPA-QUIST program (DAAD 19-01-1-0659).
F.A. has been funded by CESCA-CEPBA Large Scale Facilities
through the program "Improving the
Human Potential", contract no. HPRI-1999-CT-00071.
We are grateful for the assistance
of T. Honda with processing the samples, and for useful discussions with
A. Emperador, E. Lipparini, and Y. Tokura.

\begin{table}
\caption{
Comparison between two-dimensional CSDFT and LSDFT,
and three-dimensional LSDFT results for one single QD.
L(R) denotes the left(right) border of the MDD phase
in the $N-B_{\parallel}$ plane.
}
\begin{center}
\begin{tabular}{||c||cc||cc||cc||}
  & \multicolumn{2}{c||}{2D-CSDFT} & \multicolumn{2}{c||}{2D-LSDFT}  &
\multicolumn{2}{c||}  {3D-LSDFT}
  \\
\cline{2-7}
 $N$  &  L (T) & R (T) & L (T) & R (T) & L (T) & R (T)     \\
\hline
$\; 20 \;$   & 5.6    & 6.3    & 5.4    & 6.5   & 5.4    & 6.4    \\
 28   & 5.6    & 6.1    & 5.5    & 6.2   & 5.5    & 6.1     \\
 36   & 5.6    & 5.9    & 5.5    & 6.0   & 5.6    & 6.0     \\
\end{tabular}
\end{center}
\label{Tab1}
\end{table}

\begin{figure}
\caption[]{ (a) Schematic diagram of mesa containing
two vertically coupled quantum dots and
(b) scanning electron micrograph of a typical circular mesa.
 }
\label{fig1}
\end{figure}

\begin{figure}
\caption[]{
Double quantum well potential used in the calculations. Two electronic
densities $n(z)$ corresponding to the $N=6$ QM for $B=0$  (solid line)
and $B_{\perp}=5$ T (dashed line) are shown. $n(z)$ has been obtained
by integrating the electronic density $n(x,y,z)$ over the $x$
and $y$ coordinates. The energy of the corresponding occupied upper lying
s.p. level is also represented by a horizontal  solid line for $B=0$,
and by a horizontal dashed line for $B_{\perp}=5$ T.
}
\label{fig2}
\end{figure}

\begin{figure}
\caption[]{
Experimental $B$-field dependence of the 3rd to 6th Coulomb oscillation
peaks (g.s. electrochemical potentials for $3\le N \le 6$)
in (a) $B_{\parallel}$ case, and (b) $B_{\perp}$ case.
}
\label{fig3}
\end{figure}

\begin{figure}
\caption[]{
Theoretical ground state electrochemical potentials in the $B_{\parallel}$
case (dots) for $N\le 6$. The lines have been drawn as a to guide the
eye.
The vertical ticks along the $\mu(N)$ lines indicate phase boundaries.
The various states are identified by standard spectroscopic notation
discussed in the text.
}
\label{fig4}
\end{figure}

\begin{figure}
\caption[]{
Single particle energy levels as a function of $l$ for different
values of $B_{\parallel}$ at $N=6$. Upward(downward)
triangles denote $\uparrow$$(\downarrow)$ spin states.
Open(solid) triangles correspond to anti-bonding(bonding) states.
The horizontal lines represent the Fermi level. The value of
$B_{\parallel}$ is indicated in each panel.
}
\label{fig5}
\end{figure}

\begin{figure}
\caption[]{
Energies of the nine lower-lying non-interacting s.p.
levels as a function of $B_{\perp}$. $\Delta_{SAS}$ and $\hbar \omega$
are marked.  For each
symbol, the direction of $s_{\perp}$ is indicated in the box.
}
\label{fig6}
\end{figure}

\begin{figure}
\caption[]{
Single particle energy levels as a function of $B_{\perp}$
for $N=5$ (left panel) and 6 (right panel). For each
symbol, the direction of $s_{\perp}$ is indicated in the box.
States two fold degenerate are indicated by $\times 2$
symbol. Solid and open symbols are discussed in the text.

}
\label{fig7}
\end{figure}

\begin{figure}
\caption[]{
Theoretical ground state electrochemical potentials in the $B_{\perp}$
case
(dots) for $N \le 6$. The lines have been drawn as a to guide the eye.
The vertical marks along the $\mu(N)$ lines indicate phase
boundaries. The value of $S_{\perp}$ in each phase is
given.
We have indicated by vertical arrows downward kinks arising from s.p.
level crossings in the $N$-1 electron ground state that do not produce
phase transitions in the $N$ electron ground state.
}
\label{fig8}
\end{figure}


\begin{references}
\bibitem{nota} The frequency of the laterally confining harmonic
potential is often made $N$ dependent using as a guide the
$r$-expansion near the origin of the Coulomb potential created by a
2D uniform positive charge distribution (jellium model). This yields
$\omega \propto N^{-1/4}$.
\bibitem{box}
The simple schematic box notation in Fig. \ref{fig3} 
that represents the
molecular states is more intuitive than
the full spectroscopic notation used in Fig. \ref{fig4},
although we do not use it in Fig. \ref{fig4} because of
lack of space.
Nonetheless, it is easy to go from
one notation to the other since each box
represents a s.p. level characterized by a radial 
quantum number
$n$ and an angular momentum $l$. The black (grey) arrows
denote the thrid component of spin for electrons in bonding
(antibonding) states. At 0 T, for both bonding and antibonding
states separately, the lowest box is the $(n,l)=(0,0)\; 1s$
state, and above that are two boxes, degenerate in energy,
which are the $(0,\pm1)\; 2p$ states (see for example
$N=5$ in Fig. \ref{fig3}(a)).
For the spectroscopic notation,
the superscript $-$ of states shown in Fig. \ref{fig4}
is associated with an electron in an antibonding
box in Fig. \ref{fig3}(a), and the total $S_z$ and $L_z$
are readily obtained by adding the spin and orbital angular
momentum of all the electrons. Parity ($g$ or $u$) is
obtained by multiplying the parities of 
the s.p. states.
\bibitem{nota0} The authors of Ref. \onlinecite{Par01} have
identified the vorticity ${\cal V}$ with the external magnetic
field $\vec{B}$. While this is indeed the case for a
homogeneous 2D electron gas, and a fairly good approximation for
large QD's, this may not be a good approximation for small systems,
as in the present case. CSDFT implementations better suited
for small systems  are presented in
Refs. \onlinecite{Fer94,Pi98,Ste98,Rei99}, for example.
\bibitem{note1} Due to the convention adopted for the s.p.
wave functions,\cite{Pi01} the s.p. orbital angular momentum is
actually $-l$.
\end{references}
\end{document}